\documentclass[reprint,amsmath,amssymb,aip,jcp,superscriptaddress,twocolumn]{revtex4}
\usepackage{comment}
\usepackage{graphicx}
\usepackage{dcolumn}
\usepackage{bm}
\usepackage[acronym]{glossaries}
\usepackage{multirow}
\usepackage{hyperref}
\newacronym{CSA}{CSA}{Cartan sub-algebra}
\newacronym{LCU}{LCU}{linear combination of unitaries}

\usepackage{amsthm}

\usepackage{thmtools}
\usepackage{thm-restate}

\newcommand{\ket}[1]{|#1\rangle}
\newcommand{\bra}[1]{\langle #1|}

\newcommand{\eq}[1]{Eq.~(\ref{#1})} %
\def\be{\begin{equation}} %
\def\ee{\end{equation}} %
\newcommand{\bea}{\begin{eqnarray}}
\newcommand{\eea}{\end{eqnarray}}

\newcommand{\myceil}[1]{\lceil #1 \rceil}
\newcommand{\HH}{\hat H}

\begin{document}

\title{Reducing molecular electronic Hamiltonian simulation cost for Linear Combination of Unitaries approaches}

\author{Ignacio Loaiza}
\affiliation{Chemical Physics Theory Group, Department of Chemistry, University of Toronto, Toronto, Canada}
\affiliation{Department of Physical and Environmental Sciences, University of Toronto Scarborough, Toronto, Canada}
\author{Alireza Marefat Khah}
\affiliation{Chemical Physics Theory Group, Department of Chemistry, University of Toronto, Toronto, Canada}
\affiliation{Department of Physical and Environmental Sciences, University of Toronto Scarborough, Toronto, Canada}
\author{Nathan Wiebe}
\affiliation{Department of Computer Science, University of Toronto, Toronto, Canada}
\affiliation{Pacific Northwest National Laboratory, Richland, USA}
\author{Artur F. Izmaylov}
\affiliation{Chemical Physics Theory Group, Department of Chemistry, University of Toronto, Toronto, Canada}
\affiliation{Department of Physical and Environmental Sciences, University of Toronto Scarborough, Toronto, Canada}
\date{\today}
\begin{abstract}
We consider different \gls{LCU} decompositions for molecular electronic structure Hamiltonians. Using these \gls{LCU} decompositions for Hamiltonian simulation on a quantum computer, the main figure of merit is the 1-norm of their coefficients, which is associated with the quantum circuit complexity. It is derived that the 
lowest possible \gls{LCU} 1-norm for a given Hamiltonian is half of its spectral range. This lowest norm decomposition is practically unattainable for general Hamiltonians; therefore, multiple practical techniques to generate \gls{LCU} decompositions are proposed and assessed. A technique using symmetries to reduce the 1-norm further is also introduced. In addition to considering \gls{LCU} in the Schr\"odinger picture, we extend it to the interaction picture, which substantially further reduces the 1-norm.
\end{abstract}

\maketitle

\glsresetall

\section{Introduction}
Quantum chemistry is often regarded as one of the most promising applications for quantum computers. The number of qubits required to represent the electronic states of a molecular Hamiltonian scales linearly with the number of orbitals, whereas an exponentially large number of classical bits is required\cite{QCinQC,polynomial_dynamics}; this is commonly known as the curse of dimensionality. In addition to be able to record the quantum wavefunction one needs to be able to prepare it. 
Over the last years, several quantum algorithms for efficient eigenstate state preparation have appeared, such as Quantum Phase Estimation \cite{QPE, kitaev_QPE}. These algorithms require an implementation of the dynamical evolution operator $e^{-i\hat H t}$ and the electronic structure Hamiltonian \cite{szabo}
\begin{equation}
\hat H = \sum_{\sigma=\{\alpha,\beta\}}\sum_{ij}^N \tilde h_{ij}\hat E^{i\sigma}_{j\sigma} + \sum_{\sigma,\sigma' =\{\alpha,\beta\}}\sum_{ijkl}^N \tilde g_{ijkl}\hat E^{i\sigma}_{j\sigma}\hat E^{k\sigma'}_{l\sigma'}, \label{eq:mol_ham}
\end{equation}
where $\lbrace \sigma,\sigma' \rbrace$ are spin-$z$ projections, $\lbrace i,j,k,l\rbrace$ are spacial orbitals,
$\tilde h_{ij}$ and $\tilde g_{ijkl}$ are one- and two-electron integrals\footnote{Representing the Hamiltonian using only excitation operators is usually referred to as chemists' notation. This entails a modification to the one-electron tensor with respect to physicists' notation, which uses normal-ordered operators of the form $\hat a^\dagger_p\hat a^\dagger_q \hat a_r \hat a_s$. Our notation is related to the electronic integrals by $\tilde g_{ijkl} = \frac{1}{2}\int \int d\vec r_1 d\vec r_2 \frac{\phi_i^*(\vec r_1)\phi_j(\vec r_1)\phi_k(\vec r_2)\phi_l^*(\vec r_2)}{|\vec r_1 - \vec r_2|}$ and $\tilde h_{ij} =- \sum_k \tilde g_{ikkj} + \int d\vec r \phi_i^*(\vec r) \Big(-\frac{\nabla^2}{2} - \sum_n \frac{Z_n}{|\vec r - \vec r_n|} \Big) \phi_j(\vec r)$, with $\phi_i(\vec r)$ the one-particle electronic basis functions, and $Z_n/\vec r_n$ the charge/position of nucleus $n$.}, $N$ is the number of spacial one-electron orbitals, and   
$\hat E^p_q \equiv \hat a_p^\dagger \hat a_q$ are single excitation operators (here and for the remainder of this work, we use the indices $\lbrace p,q,r,s \rbrace$ to refer to spin-orbitals).

Implementing the time-evolution operator is a non-trivial task, known as the Hamiltonian simulation problem. This is usually done by decomposing the Hamiltonian into parts with favorable properties (e.g. fast-forwardable, unitary) \cite{pauli_simulation}, in approaches based on 1) Lie-Trotter-Suzuki formulas \cite{trotter}, 2) qubitization \cite{qubitization_df, femoco_df, THC}, 3) Dyson series \cite{LCU}, and 4) qDRIFT \cite{qdrift_1, qdrift_2, THC}. In all of these approaches, the decomposition method, together with the simulation method, define the necessary quantum resources for the simulation (i.e. number of gates and logical qubits). Some approaches, such as those based on a Lie-Trotter-Suzuki expansion, do not require additional ancilla qubits, but can require significantly deeper circuits to achieve the same worst-case error bounds for non-local Hamiltonians~\cite{childs2021theory,haah2021quantum}, whereas the tensor hypercontraction decomposition with a qubitization simulation yields the best gate provable complexity scaling to the best of our knowledge \cite{THC}.

Even though the quantum cost of realizing the time-evolution operator has a complicated dependence such as the decomposition that is being used and the particular implementation of the simulation algorithm, the gate complexity of qubitization, LCU and qDRIFT can be shown to scale with respect to the 1-norm of the Hamiltonian decomposition into a \gls{LCU} \cite{interaction, majorana_l1,LCU,time_dependent_fang,oscillatory_simulation} $\hat H = \sum_k u_k \hat U_k$, namely $\sum_k |u_k|$.  Thus minimizing this $1$-norm of coefficients is vital for reducing the costs of a quantum simulation.

In this work, we focus on ways to decompose and transform the Hamiltonian such that the resulting 1-norm is minimized. We propose three approaches that can be used for lowering the 1-norm. The first builds the \gls{LCU}'s unitaries by grouping mutually anti-commuting Pauli products in the Hamiltonian after doing the fermion to qubit mapping \cite{anticommuting}, which either maintains or lowers the 1-norm and can be obtained efficiently using a sorted-insertion algorithm\cite{SI}. The second approach shifts the Hamiltonian by using symmetry operators; efficient linear programming routines for constructing the symmetry shift are also presented. Both of these approaches are of interest for any method which decomposes the Hamiltonian as an \gls{LCU}. The third approach makes use of the interaction picture algorithm \cite{interaction} combined with a \gls{CSA} decomposition\cite{CSA}. The basic idea is to separate the Hamiltonian into two fragments: a main one with an exponential that is straightforward to calculate, and a residual with a significantly smaller 1-norm; the main fragment is obtained using a greedy optimization routine that naturally minimizes the 1-norm of the residual. The Hamiltonian simulation is then performed by rotating the residual to the interaction picture frame. Simulating the residual, as shown in Ref.~\citenum{interaction}, will generally have a significantly lower cost than the full Hamiltonian simulation. The only caveat is that the residual becomes time-dependent in this new frame, requiring a time-dependent simulation method. Since only time-independent Hamiltonians can be simulated through qubitization, we use the Dyson series approach, which can also be used for time-dependent simulations while presenting an almost optimal scaling with respect to the 1-norm and simulation time~\cite{interaction, LCU}.  It is worth noting though that other approaches such as qDRIFT that has $1$-norm scaling will also benefit from these optimizations~\cite{berry2020time}.

This rest of the paper is organized as follows. We start by reviewing the Dyson series simulation method in Sec.~\ref{subsec:Dyson}. A 1-norm lower bound theorem for \gls{LCU} decompositions is then given in Sec.~\ref{subsec:theorem}, followed by an overview of different \gls{LCU} decomposition strategies in Sec.~\ref{subsec:LCUs}. A general methodology for reducing the simulation cost by using symmetries is shown in Sec.~\ref{subsec:symmetries}. Section~\ref{subsec:interaction} reviews the essential elements of the interaction picture method \cite{interaction}, along with our procedure for partitioning the Hamiltonian into two fragments. Once all the necessary methodologies have been introduced, our numerical results along with their discussion are shown in Sec.~\ref{sec:discussion}. 

\section{Theory}
\subsection{Hamiltonian simulation by \gls{LCU} decomposition} \label{subsec:Dyson}
\begin{figure}
    \centering
    \includegraphics[width=8cm]{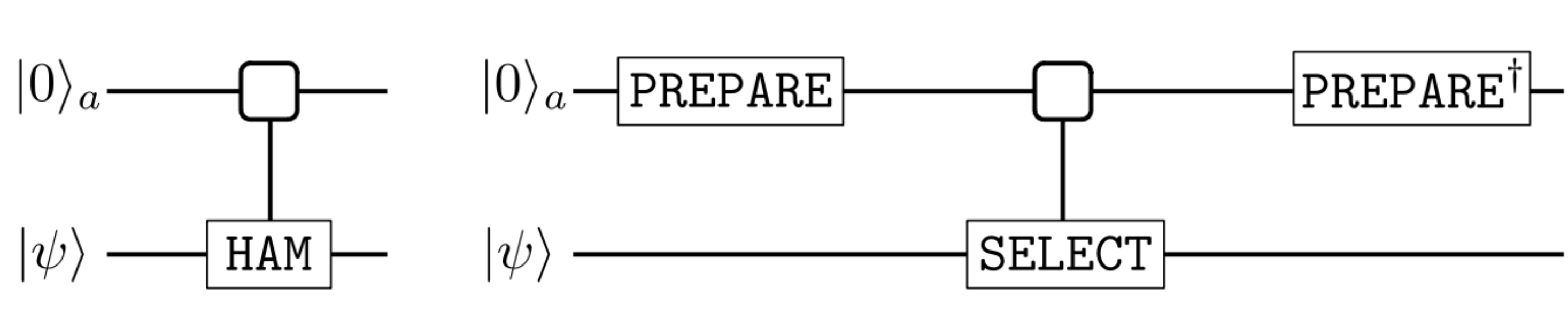}
    \caption{Arbitrary Hamiltonian oracle (left) and \gls{LCU} oracle (right) implementation by block-encoding, both with output $\ket{0}_a\otimes\frac{\hat H}{\lambda}\ket{\psi} + \ket{\textrm{remainder}}$. $\mathtt{PREPARE}\ket{0}_a = \sum_k \sqrt{\frac{u_k}{\lambda}}\ket{k}_a$ and $\mathtt{SELECT} = \sum_k \ket{k}\bra{k}_a \otimes \hat U_k$ for the \gls{LCU} shown in Eq.\eqref{eq:LCU} and $\ket{k}_a$ the $k$-th state of the ancilla register, requiring $\myceil{log_2 M}$ ancilla qubits, where $M$ is the total number of unitaries in the \gls{LCU}.}
    \label{fig:lcu_oracle}
\end{figure}

An \gls{LCU} decomposition for a Hamiltonian can be written as
\begin{equation} \label{eq:LCU}
    \hat H = \gamma \hat 1 + \sum_k u_k\hat U_k,
\end{equation}
where $u_k$'s are generally complex numbers, and $\hat U_k$'s are unitary operators.  
The identity shift can be trivially added or removed as a phase since $e^{-i(\hat H + \gamma \hat 1)t} = e^{-i\hat H t}e^{-i\gamma t \hat 1}$, while still affecting the spectral norm of the Hamiltonian. The \gls{LCU} in Eq.\eqref{eq:LCU} has an associated 1-norm for coefficients of nontrivial unitaries
\begin{equation}
    \lambda = \sum_k |u_k|.
\end{equation}
The \gls{LCU} decomposition allows one to construct a Hamiltonian oracle, a circuit that, with the usage of some ancilla qubits, is able to encode the action of $\hat H$ on an arbitrary function $\ket{\psi}$. As shown in Fig.~\ref{fig:lcu_oracle}, the output of the oracle circuit contains a spurious remainder, which is made arbitrarily small through the subsequent use of amplitude amplification. The time-evolution operator $e^{-i\hat H t}$ can always be expanded as a series: this expansion is known as a Dyson series, which reduces to a Taylor series for the case where $\hat H$ is time-independent. $e^{-i\hat H t}$ is thus realized up to an arbitrary error $\epsilon$ by implementing a truncated Dyson series using the amplitude amplified oracle circuit \cite{interaction,LCU,dyson}. 

The cost of performing the Hamiltonian simulation with an error bound $\epsilon$ scales in a linear, or nearly linear, manner with respect to the \gls{LCU}'s 1-norm \cite{interaction}, requiring
\begin{equation}
    \mathcal{O}\Big(\lambda t \ \textrm{polylog}\frac{\lambda t}{\epsilon}\Big )
\end{equation}
applications of the oracle circuit for truncated Dyson simulation methods or
\begin{equation}
    \mathcal{O}\Big(\lambda t + \ \textrm{polylog}\frac{1}{\epsilon}\Big ),
\end{equation} 
using qubitization \cite{low2019Hamiltonian,gilyen2019quantum}.  The main aim of this work is to explore different ways in which we can reduce the 1-norm $\lambda$, lowering the quantum circuit depth/complexity of the Hamiltonian simulation.

\subsection{1-norm bound} \label{subsec:theorem}
Let us start with a few general remarks on the minimum possible 1-norm and approaches for 
its minimization in practical \gls{LCU} procedures for the electronic Hamiltonian.

\begin{restatable}[1-norm bound]{theorem}{bound}
\label{thm:ck}
For a bounded hermitian operator $\hat H$, all its possible \gls{LCU} decompositions have an associated 1-norm which is lower bounded by $\Delta E/2 \equiv (E_{\rm max}-E_{\rm min})/2$, where $E_{\rm max}(E_{\rm min})$ is the highest (lowest) eigenvalue of $\hat H$.
\end{restatable}

Formal proof of the theorem is given in Appendix \ref{app:proof}. A construction that achieves the minimum 1-norm follows from the following sequence of steps: 
\begin{enumerate}
\item shift the Hamiltonian so that the largest and lowest eigenvalues become
$\pm (E_{\rm max}-E_{\rm min})/2$: 
\begin{equation}
\hat H_s = \hat H - \left(\frac{E_{\rm max}+E_{\rm min}}{2}\right) \hat 1 \label{eq:id_shift}    
\end{equation}
\item obtain a rescaled Hamiltonian $\hat H_{sr}$ such that $||\hat H_{sr}||=1$ (where $||\cdot ||$ is the spectral norm):
\begin{equation}
\hat H_{sr} = \frac{2 \hat H_s}{E_{\rm max}-E_{\rm min}} \label{eq:rescale}
\end{equation}
\item form the pair of unitaries: 
\begin{equation}
\hat U_{\pm} = \hat H_{sr}\pm i\sqrt{\hat 1-\hat H_{sr}^2}.    
\end{equation}
\end{enumerate}

This procedure gives the \gls{LCU} 
\bea\label{eq:HLCU}
\hat H = \left(E_{\rm min}+\frac{\Delta E}{2}\right) \hat 1 + \frac{\Delta E}{4} \left(\hat U_{+}+\hat U_{-}\right),
\eea
with $\Delta E/2$ its corresponding 1-norm.

The main problem of achieving this lowest 1-norm is the practical construction of $\hat U_{\pm}$; for most molecular Hamiltonians they are highly multi-particle operators (i.e. including up to $2N$-electron excitations) because of the square-root in their definitions, and their calculation/implementation requires a complicated quantum singular value transformation to implement~\cite{gilyen2019quantum} the square-roots in an interval $\|\hat{H}_{sr}\| \in [0,1-\delta]$ for some $\delta>0$. Still, this theorem gives us a bound on the best achievable value for the 1-norm, which is useful to assess the quality of \gls{LCU} decompositions.

\subsection{Unitary operators for \gls{LCU} decompositions} \label{subsec:LCUs}

In what follows we consider various practical forms of unitaries $\hat U_k$ that appear when one is constructing an \gls{LCU} representation for the electronic structure Hamiltonian [\eq{eq:mol_ham}].

\subsubsection{Pauli products}
The simplest \gls{LCU} decomposition uses the qubit representation of the Hamiltonian after it has been transformed from its fermionic representation by using one of the many available mappings (e.g. Jordan-Wigner and Bravyi-Kitaev transformations \cite{jordan_wigner,bk1,bk2,bk3,trans_comparison}):
\begin{equation} \label{eq:pauli}
    \hat H = \sum_{k=1}^{N_p} c_k \hat P_k,
\end{equation}
where $\hat P_k$ are products of Pauli operators on different qubits, $c_k$ are constants, and $N_p$ is the total number of Pauli products. Since Pauli products are already unitary operators, Eq.\eqref{eq:pauli} defines an \gls{LCU} decomposition where the 1-norm is given by
\begin{equation}
    \lambda^{(\textrm{P})} = \sum_k |c_k|. 
\end{equation}
$\lambda^{(\textrm{P})}$ can be related to electron integrals $\tilde h_{ij}$ and $\tilde g_{ijkl}$ as 
\begin{align}
    \lambda^{(\textrm{P})} &= \sum_{ij} \Big\lvert \tilde h_{ij} + 2\sum_k \tilde g_{ijkk}\Big\rvert + \sum_{i>k,j>l} |\tilde g_{ijkl} - \tilde g_{ilkj}| \nonumber \\
    &\ \ \ \ \ + \frac{1}{2}\sum_{ijkl}|\tilde g_{ijkl}|. \label{eq:majorana_l1}
\end{align}
This relation is obtained using the Majorana representation for the Hamiltonian (detailed in Appendix~\ref{app:tensors}). \\
 
One way to improve $\lambda^{(\textrm{P})}$ is to apply orbital transformations $\hat U_O$ 
\begin{equation}\label{eq:orb_rot}
    \hat U_O(\vec \theta) = e^{\sum_{i > j} \theta_{ij}\sum_\sigma(\hat E^{i\sigma}_{j\sigma} - \hat E^{j\sigma}_{i\sigma})}
\end{equation}
before transforming the Hamiltonian to qubit or Majorana representation~\cite{majorana_l1}, 
 $\hat U_O\hat H \hat U_O^\dagger$. 
 One can optimize the generator coefficients $\theta_{ij}$ of orbital rotations in $\hat U_O(\vec \theta)$ with the 1-norm in Eq.\eqref{eq:majorana_l1} as the cost function.

\subsubsection{Grouping of anti-commuting Pauli products}
Another way of reducing 1-norm of the Pauli product decomposition of the Hamiltonian [\eq{eq:pauli}] 
is via grouping of anti-commuting Pauli products\cite{anticommuting} 
\begin{equation}\label{eq:ACg}
    \hat H = \sum_n a_n \hat A_n,
\end{equation}
where 
\bea
a_n &=& \sqrt{\sum_{i\in K_n} |c_i|^2}, \\
\hat A_n &=& \frac{1}{a_n} \sum_{k\in K_n} c_k \hat P_k,
\eea
$\hat A_n$ are linear combinations of anti-commuting Pauli products, 
$\lbrace\hat P_{k},\hat P_{k'}\rbrace=2\delta_{kk'}\hat 1 $, and 
$\lbrace K_n\rbrace$ are groups of corresponding indices. 

$\hat A_n$ are larger unitary transformations, and it can be shown that
1-norm of coefficients in \eq{eq:ACg} is always lower or equal than the 1-norm of the initial LCU decomposition:
comparison of 1-norms for decompositions in Eqs.~\eqref{eq:pauli} and \eqref{eq:ACg} 
\bea
    \lambda^{(\textrm{AC})} &=& \sum_n |a_n| = \sum_n \sqrt{\sum_{i\in K_n} |c_i|^2}\\
    \lambda^{(\textrm{P})} &=& \sum_k |c_k| = \sum_n \sum_{i\in K_n} |c_i|
\eea
and using the triangle inequality for each $K_n$,
\begin{equation} \label{eq:l1_ineq}
    \sum_{i\in K_n} |c_i| \geq \sqrt{\sum_{i\in K_n} |c_i|^2},
\end{equation}
proves that $\lambda^{(\textrm{AC})} \leq \lambda^{(\textrm{P})}$, where the equality takes place only 
for the trivial case where no grouping is made.
Further note that from the Cauchy-Schwarz inequality
\begin{equation}
    \sum_{i\in K_n} |c_i| \le \sqrt{|K_n|} \sqrt{\sum_{i\in K_n} |c_i|^2},
\end{equation}
the gulf between these two bounds can be as much as a factor of $\sqrt{|K_n|}$ suggesting that a substantial improvement can be attained in cases where the typical values of $|K_n|$ seen are large.

 Once the anti-commuting groups have been found, each unitary $\hat A_n$ can be implemented as a product of unitaries:
\begin{equation}
    \hat A_n = \prod_{k\in K_n \uparrow} e^{i \theta_k \hat P_k} \prod_{k\in K_n \downarrow} e^{i \theta_k \hat P_k}, 
\end{equation}
where the $\uparrow$($\downarrow$) stands for ascending(descending) through $K_n$, and 
\begin{equation}
    \theta_k = \frac{1}{2}\textrm{arcsin}\frac{c_k}{\sqrt{\sum_{\substack{i \in K_n \\ i\leq k}} c_i^2}}.
\end{equation}

\subsubsection{Fermionic reflections} 
\label{subsubsec:fermionic}

One can use fermionic operator algebra to construct an LCU decomposition. 
It is convenient to start with factorized decomposition of one- and two-electron 
terms of $\HH$ [\eq{eq:mol_ham}] introduced in double factorization (DF) method\cite{qubitization_df, femoco_df} 
and the Cartan subalgebra approach (CSA)\cite{CSA} 
\bea
\HH_1 &=&  \sum_{\sigma=\{\alpha,\beta\}}\sum_{ij}^N \tilde h_{ij}\hat E^{i\sigma}_{j\sigma} \\
&=& \hat U_1 \Big(\sum_{i\sigma} \lambda_{ii}^{(1)} \hat n_{i\sigma} \Big) \hat U_1^\dagger \\
\HH_2 &=& \sum_{\sigma,\sigma' =\{\alpha,\beta\}}\sum_{ijkl}^N \tilde g_{ijkl}\hat E^{i\sigma}_{j\sigma}\hat E^{k\sigma'}_{l\sigma'} \\
&=& \sum_{m=2}^{M} \hat U_m \left( \sum_{ij,\sigma \sigma'} \lambda_{ij}^{(m)} \hat n_{i\sigma} \hat n_{j\sigma'} \right) \hat U_m^\dagger,
\eea
where $\hat U_m$ are orbital rotations as in \eq{eq:orb_rot}, 
$\hat n_p \equiv \hat E^p_p$ are the occupation number operators, and $\lambda_{ij}^{(m)}$ are parameters 
of the decomposition. The difference between DF and CSA decompositions is in the rank 
of $\lambda_{ij}^{(m)},~m>1$ matrices, DF has $\lambda_{ij}^{(m)} = \epsilon_{i}^{(m)}\epsilon_{j}^{(m)}$ as 
an outer product of $\epsilon_{i}^{(m)}$ vector, which makes ${\rm rank}(\lambda^{(m)})=1$, while CSA 
does not have any restriction on the $\lambda_{ij}^{(m)}$ rank. $\HH_2$ decompositions are truncated 
for fragments whose $\lambda_{ij}^{(m)}$ are smaller than some threshold. 

 The convenience of these decompositions are in simplicity of transforming $\hat n_p$ operators into reflections,\cite{femoco_df, qubitization_df, THC} $\hat n_p \rightarrow (2\hat n_p - \hat 1) \equiv \hat r_p$, where $\hat r_p$ is the reflection on spin-orbital $p$. Accounting for idempotency and hermiticity of $\hat n_p$, one can easily check that 
 $\hat r_p^2 = 1$ and $\hat r_p^\dagger = \hat r_p$. Another important property of $\hat n_p$'s is their 
 commutativity that allows one to transform the two-electron fragment in LCU  
 \bea\notag
 \HH_2\rightarrow\HH_2^{\rm (LCU)} &=& 
  \sum_{\sigma \sigma'}\sum_m \hat U_m \left(\sum_{ij}\frac{\lambda_{ij}^{(m)}}{4} 
  \hat r_{i\sigma} \hat r_{j\sigma'} \right) \hat U_m^\dagger.
 \eea
  To use $\HH_2^{\rm (LCU)}$ as a part of the full Hamiltonian, the one-electron part requires the following adjustment
\begin{equation} \label{eq:Tmod}
    \hat H_1' = \hat H_1 + 2\sum_{\sigma}\sum_{ijk} \tilde g_{ijkk} \hat E^{i\sigma}_{j\sigma}.
\end{equation}

 The modified one-electron operator can be decomposed as
\begin{equation} \label{eq:Tdiag}
    \hat H_1' = \hat U_1 \left(\sum_{\sigma,i} \mu_i \hat n_{i\sigma} \right)\hat U_1^\dagger,
\end{equation}
where $\mu_i$ are new parameters. 
Switching $\hat n_p \rightarrow \hat r_p$ yields up to a constant shift 
\begin{equation} \label{eq:ob_lcu}
 \HH_1'\rightarrow \hat H_1^{\rm (LCU)} = \hat U_1 \left(\sum_{\sigma,i} \frac{\mu_i}{2} \hat r_{i\sigma} \right)\hat U_1^\dagger.
\end{equation}
Thus, we have 1-norm $\lambda^{(\textrm{F})}= \lambda_1 + \lambda_2$ associated with this LCU procedure,
where $\lambda_1$ ($\lambda_2$) is 1-norm of $\hat H_1^{\rm (LCU)}$ ($\hat H_2^{\rm (LCU)}$)
\begin{align}
    \lambda_1 &= \sum_i |\mu_i|, \label{eq:l1_T} \\
    \lambda_2 &= \frac{1}{4}\sum_m \sum_{\substack{\sigma \sigma', ij\\ i\sigma \neq j\sigma'}}|\lambda_{ij}^{(m)}| \nonumber \\
    &\quad = \sum_m\sum_{ij} |\lambda_{ij}^{(m)}| - \frac{1}{2} \sum_m \sum_i |\lambda_{ii}^{(m)}|, \label{eq:l1_V}
\end{align}
Note that the obtained fermionic LCU decomposition can be transformed to a qubit LCU form 
using any fermion-qubit mappings. 

\subsubsection{Square-root unitarization technique} 
\label{subsubsec:sr}

Another way to transform the quadratic polynomials 
\begin{equation} \label{eq:ferm_frag}
    \hat f^{(m)} = \sum_{ij,\sigma\sigma'}\lambda_{ij}^{(m)}\hat n_{i\sigma}
    \hat n_{j\sigma'} 
\end{equation}
into unitaries is via steps described in Theorem 1. This procedure gives the lowest 1-norm for LCU of each fragment. First, we shift and rescale each fragment so that their eigenvalues are in the interval $[-1,1]$ (i.e. $\hat f^{(m)} \rightarrow \hat f^{(m)}_{sr}$, as seen in Eqs.(\ref{eq:id_shift},\ref{eq:rescale})). Second, unitary operators can be defined for each fragment as
\begin{equation}
    \hat V_m^{(\pm)} = \hat f^{(m)}_{sr} \pm i\sqrt{\hat 1 - (\hat f^{(m)}_{sr})^2}.
\end{equation}
This leads to the LCU of the two-electron Hamiltonian up to a constant as
\bea
\HH_2^{\rm (LCU)} = \sum_{m=2}^M \frac{ ||\hat f^{(m)}||_\Delta}{2} \hat U_m (\hat V_m^{(+)} + \hat V_m^{(-)}) \hat U_m^\dagger, 
\eea 
where $||\hat f^{(m)}||_\Delta\equiv f^{(m)}_{\max} - f^{(m)}_{\min}$ is a spectral range of $\hat f^{(m)}$ and 
$f^{(m)}_{\max(\min)}$ is the maximum (minimum) eigenvalue of fragment $\hat f^{(m)}$.
The corresponding 1-norm is
\begin{equation}
    \lambda_2^{(sr)} = \frac{1}{2}\sum_m ||\hat f^{(m)}||_\Delta,
\end{equation}

In general, the explicit construction and implementation of $\hat V_m^{(\pm)}$ unitaries 
becomes prohibitively expensive as the number of orbitals grows, as the square-root function 
includes all possible products of the occupation numbers. 
Using a different scaling such that $\|\hat f^{(m)}_{sr}\|\le  1-\delta$, this unitary operator 
can be approximated within error $\epsilon$ using $O((\frac{1}{\delta})\log(1/\epsilon)) $ 
invocations of the fragment $\hat{f}^{(m)}$~\cite{gilyen2019quantum}. 
Nevertheless, given the complexity of implementing such circuits, the square-root unitarization is primarily used as a lower bound of the 1-norm for fermionic decompositions. 

 This procedure can also be done for the one-electron term $\hat H_1'$ (Eq.\eqref{eq:Tdiag}), obtaining
\begin{equation} \label{eq:sqrt_l1T}
    \lambda_1^{(sr)} = \Big(\sum_{\mu_i>0} \mu_i - \sum_{\mu_i<0} \mu_i \Big) = \sum_i |\mu_i|.
\end{equation}
We note that this 1-norm is the same as that in Eq.\eqref{eq:l1_T}. This shows that, 
using the fermionic LCU decomposition in \eq{eq:Tdiag} gives the LCU with the lowest possible 1-norm for the 
one-electron part since its 1-norm coincides with that in the square-root approach. 

\subsubsection{Chebyshev complete-square block-encoding:} 

The relation $\lambda_{ij}^{(m)} = \epsilon_i^{(m)}\epsilon_j^{(m)}$ in the double factorization approach allows one
to factorize quadratic polynomials into complete squares 
\bea\label{eq:cs}
\hat f^{(m)} = \sum_{ij,\sigma \sigma'} \lambda_{ij}^{(m)} \hat n_{i\sigma} \hat n_{j\sigma'} = 
\left(\sum_{i,\sigma} \epsilon_i^{(m)} \hat n_{i\sigma} \right)^2.
\eea
Given a fragment with a complete-square structure $\hat f = \hat l^2$, its implementation can be efficiently done in two steps. First, we transform $\hat l$ into an \gls{LCU} by the mapping $\hat n_p \rightarrow \hat r_p$:
\begin{equation} \label{eq:l}
    \hat l = \left(\sum_i \epsilon_i\right) \hat 1 + \sum_i \frac{\epsilon_i}{2} \sum_\sigma \hat r_{i\sigma}.
\end{equation}
The corresponding 1-norm of this \gls{LCU} becomes $\sum_i |\epsilon_i|\equiv \lambda_l$, which as shown by Eq.\eqref{eq:sqrt_l1T} is optimal. Second, the qubitization technique  \cite{low2019Hamiltonian,qubitization_df} is used for block-encoding $\hat f$. This is done by implementing the Chebyshev polynomial $T_2[\hat l_n] = 2\hat l_n^2 - \hat 1$, with $\hat l_n \equiv \frac{\hat l}{\lambda_l}$. Thus, by presenting
\begin{equation} \label{eq:cheby}
    \hat f = \frac{\lambda_l^2}{2}T_2[\hat l_n] + \frac{\lambda_l^2}{2}\hat 1,
\end{equation}
this fragment can be implemented with an associated 1-norm cost of $\frac{\lambda_l^2}{2}$. We will refer to this technique as a complete-square encoding. Note that the implementation of $\hat l$ as an \gls{LCU} is done without the constant shift appearing in Eq.\eqref{eq:l}. This shift appears as a one-body correction when implementing $\hat f$ through the complete-square encoding. As demonstrated in Ref.\citenum{THC}, the adjusted one-electron Hamiltonian in Eq.\eqref{eq:Tmod} can be recovered by adding the one-body corrections from all fragments in the two-electron Hamiltonian $\hat H_2$. Constant terms must also be accounted for and can be trivially added after representing $\hat H$ as an \gls{LCU}.

Finally, we note that the 1-norm obtained for $\hat f$ through the square-root decomposition is the same as that obtained through the complete-square encoding. The 1-norm for the two-body part of $\hat f^{(m)}$ becomes
\begin{equation} \label{eq:sr_cost}
    \lambda_2^{(m)} = \frac{1}{2}\left(\sum_i \lvert\epsilon_i^{(m)}\rvert \right)^2.
\end{equation}

Thus, complete-square fragments can be encoded optimally with the lowest possible 1-norm, while for the more general case of $\lambda^{(m)}_{ij}\neq \epsilon^{(m)}_i\epsilon^{(m)}_j$ either the square-root technique can be applied or their $\lambda^{(m)}_{ij}$ can be decomposed into a sum over complete squares.

\subsection{Norm reduction by using symmetry shifts} 
\label{subsec:symmetries}

The idea of symmetry shifts $\hat S$ uses the same rationale as that of shifting the Hamiltonian by $\gamma \hat 1$: the overall 1-norm of the \gls{LCU} decomposition can be lowered by decomposing $\hat H- \hat S$ instead of $\hat H$. By considering an arbitrary symmetry operator $\hat S$ for which $[\hat H,\hat S] = 0$, the time-evolution operator becomes
\begin{equation}
    e^{-i\hat H t} = e^{-i\hat S t} e^{-i (\hat H - \hat S) t}.
\end{equation}
For a wavefunction that obeys symmetry constraints, $e^{-i\hat S t}$ only introduces a phase, making the simulation problem equivalent to the simulation of $\hat H - \hat S$. In this section we focus on ways to find this shift operator such that the 1-norm of the resulting \gls{LCU} is lowered. Note that this symmetry shift technique can also improve Trotter methods as well.

$\hat S$ can be built by an arbitrary function $f(\{\hat S_u\})$ over a set of symmetries $\mathbb{S}=\{\hat S_u\}$ satisfying the conditions $[\hat S_u,\hat S_v]=[\hat S_u, \hat H]=0,\ \forall u,v \in 1,...,|\mathbb{S}|$. The symmetries $\mathbb{S}$ can be selected from existing molecular symmetries, such as the electron number operator $\hat N_e$, the z-projection of electron spin $\hat S_z$, the total electronic spin $\hat S^2$, and molecular point group symmetries. However, in this study we restrict the pool of symmetries to $\mathbb{S} = \{\hat N_e\}$. This choice preserves the Hamiltonian structure in Eq.\eqref{eq:mol_ham} where the fermionic tensors $\tilde h_{ij}$ and $\tilde g_{ijkl}$ run over spacial orbitals instead of spin-orbitals. There are three benefits to this structure: (1) all \gls{LCU} methodologies can be used without modification, (2) it allows for efficient compilation of \gls{LCU} unitaries on a quantum computer\cite{THC}, and (3) it improves the efficiency of storage and manipulation of these tensors on a classical computer.

The symmetry shift procedure contains two steps. First, the optimal shift for the two-electron operator is found, such that the 1-norm of $\hat H_2(s_2) \equiv \hat H_2 - s_2 \hat N_e^2$ is minimized. Seconds, the modified one-body Hamiltonian $\hat H_1'(s_2)$ is obtained [\eq{eq:Tmod}] and the 1-norm of $\hat H_1'(s_2) - s_1 \hat N_e$ is minimized, yielding $\hat S = s_1 \hat N_e + s_2 \hat N_e^2$. This procedure reduces 1-norm of $\hat H - \hat S$, while preserving the $\hat H$ structure so that all \gls{LCU} methodologies can be applied. Efficient routines for finding $s_1$ and $s_2$ are detailed in Appendix~\ref{app:sym_L1}.

\subsection{Hamiltonian simulation in the interaction picture} 
\label{subsec:interaction}
Here, we consider the interaction picture Hamiltonian simulation introduced in Ref.~\citenum{interaction}. 
The Hamiltonian is split in two fragments, $\HH=\HH_0+\HH_R$, where $\HH_0$ is exactly solvable or fast-forwardable and $\hat H_R$ is a residual part. It is beneficial to achieve $||\HH_0||\gg||\hat H_R||$ in this splitting. 
The simulation of the molecular Hamiltonian then requires the time-dependent simulation of the residual Hamiltonian
in the interaction picture, which reduces the overall cost of the simulation.  \\

For selecting $\HH_0$ we use the CSA approach, 
by finding the orbital rotation amplitudes $\vec \theta$ and occupation number 
parameters $(\vec \mu,\vec \lambda)$ in
\begin{equation}
    \HH_0= \hat U(\vec \theta)\Big(\sum_{\sigma,i} \mu_i \hat n_{i\sigma} + \sum_{\sigma\sigma',i>j} \lambda_{ij}\hat n_{i\sigma}\hat n_{j\sigma'} \Big) \hat U^\dagger(\vec \theta),
\end{equation}
 to minimize $||\hat H - \HH_0||$, where the norm is defined as a 
 sum of $L_2$ norms for one- and two-electron tensors.
 $\HH_0$ corresponds to a largest mean-field solvable part\cite{meanfield} of $\hat H$. 

 In the interaction picture, the time-dependent Schrödinger equation is
 \begin{equation}
     i\partial_t \ket{\psi_I(t)} = \hat H_R^{(I)}(t) \ket{\psi_I(t)},
 \end{equation}
 where $\ket{\psi_I(t)} \equiv e^{i\HH_0t} \ket{\psi(t)}$ and 
 \begin{equation} \label{eq:interaction_tdse}
    \hat H_R^{(I)}(t) = e^{i\HH_0t} \hat H_R e^{-i\HH_0t}.
 \end{equation}
 Simulating the evolution under the full Hamiltonian is thus equivalent to rotating to the interaction frame by 
 $e^{i\HH_0t}$, doing the simulation with the time-dependent residual $\hat H_R^{(I)}(t)$, and rotating back to the Schrödinger frame by $e^{-i\HH_0t}$.
 
 As shown in Ref.~\citenum{interaction}, the associated simulation cost for the interaction picture method using a truncated Dyson series is 
 \begin{equation}
    \mathcal{O}\left(\lambda_R t \ \textrm{polylog}\left(\frac{\lambda t}{\epsilon}\right)\right )
\end{equation}
applications of the $\hat H_R$ Hamiltonian oracle and of $e^{i\tau \HH_0}$ (with different values of $\tau)$, where $\lambda_R$ is the 1-norm of the \gls{LCU} decomposition of $\hat H_R$. 

\section{Results and Discussion} 
\label{sec:discussion}
\begin{table*}[]
\centering
\begin{tabular}{|c|c|c|c|c|c|c|c|c|c|}
\hline
System                               & Symmetry shift & $\Delta E/2$  & Pauli        & OO-Pauli      & AC           & OO-AC        & DF         & GCSA-F        & GCSA-SR       \\ \hline
\multirow{2}{*}{H$_2$}               & --             & 0.82          & 1.58(4)      & 1.58(4)       & 1.49(4)      & 1.49(4)      & 1.37(2)       & 1.77(4)       & 1.37(2)       \\
                                     & yes            & 0.66          & 0.84(4)      & 0.84(4)       & 0.79(5)       & 0.79(5)      & 0.75(2)      & 0.84(4)       & 0.74(2)       \\ \hline
\multirow{2}{*}{LiH}                 & --             & 4.93          & 13.0(10)     & 12.4(11)      & 10.2(7)       & 10.2(8)     & 9.34(5)       & 11.0(11)      & 8.25(5)       \\
                                     & yes            & 3.57          & 7.62(10)     & 7.02(10)      & 5.13(7)       & 5.03(7)     & 4.76(5)       & 5.48(11)      & 4.61(5)       \\ \hline
\multirow{2}{*}{BeH$_2$}             & --             & 9.99          & 22.8(10)     & 21.9(12)      & 18.0(8)       & 17.9(10)     & 16.4(5)       & 20.1(12)      & 14.6(5)       \\
                                     & yes            & 7.31          & 14.2(10)     & 13.0(12)      & 10.2(8)       & 9.83(8)     & 9.77(5)       & 11.5(11)      & 9.58(5)       \\ \hline
\multirow{2}{*}{H$_2$O}              & --             & 41.9          & 71.9(11)     & 60.1(12)      & 57.2(8)       & 55.7(9)     & 53.7(5)       & 59.2(12)      & 50.6(5)       \\
                                     & yes            & 28.9          & 46.0(11)     & 37.7(12)      & 34.4(8)       & 32.9(9)     & 32.7(5)       & 36.1(12)      & 31.9(5)       \\ \hline
\multirow{2}{*}{NH$_3$}              & --             & 33.8          & 68.6(12)     & 54.5(12)      & 48.8(9)       & 46.8(12)     & 44.7(6)       & 50.6(13)      & 40.6(6)       \\
                                     & yes            & 23.1          & 46.3(12)     & 34.6(12)      & 29.8(9)       & 27.8(12)     & 28.1(6)       & 31.8(13)      & 26.5(6)       \\ \hline \hline
\multirow{2}{*}{Linear fit slope} & --    & 0.53$\pm$0.03        & 1      & 0.83$\pm$0.02 & 0.76$\pm$0.02 & 0.73$\pm$0.02 & \textbf{0.70$\pm$0.02} & 0.79$\pm$0.02 & 0.65$\pm$0.03 \\
                                              & yes   & 0.37$\pm$0.02 & 0.65$\pm$0.01 & 0.52$\pm$0.01 & 0.46$\pm$0.01 & \textbf{0.43$\pm$0.01} & \textbf{0.43$\pm$0.01} & 0.48$\pm$0.01 & 0.42$\pm$0.01 \\ \hline

\end{tabular}
\caption{1-norms for molecular Hamiltonian using different LCU decompositions: 
$\Delta E/2 = (E_{\rm max}-E_{\rm min})/2$ is the lowest 1-norm possible; Pauli, Pauli products; AC, anti-commuting Pauli product grouping; OO-Pauli, Pauli products with orbital optimization for 1-norm;  AC-OO, orbital optimization scheme with subsequent anti-commuting Pauli product grouping; DF, double factorization; GCSA, Greedy CSA decomposition. Symmetry shift procedure is outlined in Sec.~\ref{subsec:symmetries}. Suffixes -F and -SR for GCSA correspond to fermionic unitarization and square-root unitarization techniques, respectively. DF results correspond to complete-square technique, which yield a minimal 1-norm. 
Values in parenthesis represent $\myceil{\log_2(\textrm{\#\ unitaries})}$, which is associated with the necessary number of ancilla qubits and circuit depth (see Refs.\citenum{THC,femoco_df} for a more detailed discussion). A cut-off threshold of $10^{-6}$ was used for counting unitaries in fermionic unitarizations. The linear fit was obtained by associating to each molecule two coordinates: $x$ is 
the 1-norm of the Pauli product LCU, and $y$ is the 1-norm of the considered LCU method.  The slope of the 
linear regression is given with the associated standard error ($\pm$). Highlights show methods with best scaling that are viable to implement on a quantum computer.}
\label{tab:results_1}
\end{table*}

\begin{table*}[]
\centering
\begin{tabular}{|c|c|c|c|c|c|c|c|c|}
\hline
System                       & $\Delta E/2$    & Pauli       & OO-Pauli    & AC          & OO-AC      & DF           & GCSA-F      & GCSA-SR      \\ \hline
H$_2$                        & 0.20            & 0.30(3)     & 0.30(3)     & 0.30(3)     & 0.30(3)    & 0.20(1)      & 0.30(3)     & 0.20(1)      \\ \hline
LiH                          & 0.80            & 3.13(10)    & 2.88(10)    & 1.50(7)     & 1.49(7)    & 1.40(5)      & 2.12(11)    & 1.53(5)      \\ \hline
BeH$_2$                      & 1.00            & 5.78(10)    & 4.41(10)    & 2.60(8)     & 2.33(8)    & 2.89(5)      & 3.95(12)    & 2.44(5)       \\ \hline
H$_2$O                       & 2.38            & 9.18(11)    & 7.77(11)    & 4.32(8)     & 3.91(8)    & 4.47(5)      & 7.94(12)    & 5.34(6)      \\ \hline
NH$_3$                       & 3.01            & 14.3(13)    & 11.2(13)    & 5.86(10)    & 5.23(10)   & 6.09(6)      & 10.2(13)    & 6.42(6)       \\ \hline \hline
Linear fit slope & 0.039$\pm$0.003 & 0.172$\pm$0.022 & 0.139$\pm$0.016 & 0.075$\pm$0.008 & \textbf{0.067$\pm$0.007} & 0.078$\pm$0.009 & 0.131$\pm$0.085 & 0.085$\pm$0.006 \\ \hline
\end{tabular}
\caption{Same as Table~\ref{tab:results_1}, but for residual Hamiltonian $\hat H_R$. The linear fit is calculated with respect to the Pauli product LCU for $\hat H$.}
\label{tab:results_2}
\end{table*}

\subsection{Schr\"odinger picture}
Table~\ref{tab:results_1} shows the 1-norms obtained for different methods on molecular Hamiltonians. A comparison of each method was made with respect to the simple Pauli product decomposition. 
The linear fit for each LCU decomposition was done by associating to each molecule two coordinates: 
$x$ is the 1-norm of the Pauli product LCU, and $y$ is the 1-norm of the considered LCU method. 
The slope of the linear fit quantifies the average 1-norm reduction of each method 
with respect to the Pauli product LCU decomposition. Note that here we do not obtain the scaling of 1-norm with respect to the system size. Such a scaling consideration would require a more systematic approach to increasing the system size and is left for future studies.

For the qubit-based methods, the orbital optimization scheme presents an improvement with respect to the Pauli product LCU, as expected from the analysis in Ref.~\citenum{majorana_l1}. This trend is also true for the anti-commuting grouping technique, and the most significant improvement was obtained when combining both techniques.
Symmetry shift presented a significant improvement in 1-norm for all systems. It is the best qubit-based method when combined with the anti-commuting grouping and orbital optimization techniques, with an average decrease of 1-norm by $57\%$. \\

Before discussing the fermionic methods, we would like to emphasize that complete-square encoding will yield the lowest possible 1-norm for fragments with a complete-square structure. Therefore, the optimal 1-norms also correspond to those of DF with complete-square encoding (used in Refs.~\citenum{femoco_df,THC}). We also note that implementing fragments as a single unitary not only reduces the 1-norm, but also the number of unitaries and consequently the number of ancilla qubits. Thus, the square-root technique works as a lower bound for the 1-norm and can be used to know how much the fermionic LCU can be further improved along with providing intuition about the minimal number of ancillary qubits needed. Since the number of ancilla qubits will also depend on the particular implementation of the controlled unitaries \cite{THC,femoco_df}, Table~\ref{tab:results_1} shows $\myceil{\log_2 M}$, where $M$ is the number of unitaries in the \gls{LCU} and the number of ancillas scales with $\mathcal{O} (\myceil{\log_2 M})$. \\ 

As shown in Table~\ref{tab:results_1}, all fermionic methods show a significant improvement with application of symmetry shift. The symmetry-shifted greedy CSA approach with the square-root technique (GCSA-SR) gave lowest 1-norms, with an average reduction of $58\%$. However, this is an unattainable lower bound. Thus, we highlight methods with the best average gain that use viable implementation methods by employing simple quantum circuits. Before symmetry shift, the best fermionic method is DF, having an average decrease of the 1-norm by $30\%$. When the symmetry shift was applied, all methods presented a significant improvement, having the largest reduction of $57\%$ for DF.

Even though the CSA decomposition is a more general ansatz than DF, the greedy algorithm used in CSA for finding the fragments is heuristic and sacrifices some flexibility of a full optimization in exchange for computational efficiency. This makes the number of fragments and as a result the number of ancilla qubits appearing in GCSA larger than those in DF. 

It should also be noted that doing the GCSA decomposition is a computationally expensive process compared to DF; these methods respectively require a non-linear optimization and a Cholesky decomposition. When considering the symmetry shift, both anti-commuting with orbital optimization and DF with a fermionic unitarization methods present the best improvement for 1-norm while having the lowest classical pre-processing cost. Both these methods show an improvement of $57\%$, offering a significant improvement over all methods without symmetry shift. 

 Although both DF and CSA decompositions could be done over the full Hamiltonian by representing 
 it as a single two-electron operator (see Appendix~\ref{app:tensors}), separating the one- and two-electron terms yields a very significant improvement of the 1-norm, while also allowing for manipulations to be done directly on spacial orbitals (as opposed to spin-orbitals), which greatly lowers the classical pre-processing cost of the decompositions.

Finally, we discuss how our results compare to the 1-norm lower bound $\Delta E/2$. 
When not considering symmetry shift, DF lowered the 1-norm by $30\%$ of the possible reduction, 
while with symmetry shift it captured $57\%$. When dealing with symmetry shifted molecular Hamiltonians, 
both DF and the orbital-optimized anti-commuting grouping obtain a 1-norm that is very close to the lower bound along a small number of required ancilla qubits, placing them as almost optimal decomposition methods. 

\subsection{Interaction picture}
Table~\ref{tab:results_2} provides results for the same LCU method but in the interaction picture. 
The linear fit results are also obtained by comparing the 1-norms with respect to the Pauli product LCU 1-norm of the molecular Hamiltonian in the Schr\"odinger picture. \\

Here, performing the symmetry shift does not yield any significant improvements on any of the methods and thus 
is not shown in Table~\ref{tab:results_2}. This can be attributed to an implicit inclusion of any symmetry shift of $\HH_0$ in its $\mu_i$ and $\lambda_{ij}$ coefficients: $\HH_0 - \sum_u s_u\hat S_u \equiv \hat U(\vec \theta) \Big( \sum_{\sigma,i}\tilde \mu_i(\vec s) \hat n_{i\sigma} + \sum_{\sigma\sigma',i>j} \tilde \lambda_{ij} (\vec s) \hat n_{i\sigma}\hat n_{j\sigma'} \Big) \hat U^\dagger(\vec \theta)$ since $[\hat S_u,\hat U(\vec \theta)]=0$ and $\hat S_u$ is a polynomial of occupation numbers.
The DF technique is the most optimal fermionic method, with an average 1-norm decrease of $92\%$ with 
respect to the Pauli product LCU of the molecular Hamiltonian. Surprisingly, the results from GCSA are not consistently better than those of DF. We attribute this to the heuristic nature of the greedy optimization. DF is thus a better alternative than GCSA for decomposing $\hat H_R$, presenting a greater 1-norm reduction, smaller number of fragments, and a lower classical pre-processing cost. \\

For the qubit-based methods, the anti-commuting grouping presented the greatest 1-norm reduction. We observed that applying the orbital optimization scheme further improved the results when combined with the anti-commuting grouping. The combination of these techniques yielded an average 1-norm reduction of $93\%$, while without the orbital optimization scheme the anti-commuting grouping presented an average decrease of $92\%$. 

\section{Conclusion}

We have presented a wide set of techniques for performing the \gls{LCU} decomposition 
of molecular Hamiltonians, using the associated 1-norm of the decomposition as the main figure 
of merit. The greatest 1-norm reduction was observed for the interaction picture methodology 
combined with the orbital optimization scheme and anti-commuting Pauli products grouping, with an average 1-norm decrease of $93\%$ with respect to the Pauli product LCU of the molecular Hamiltonian $\hat H$. 
All methods that work with the residual $\hat H_R$ of the interaction picture greatly improved the performance over decomposing $\hat H$, decreasing the 1-norm by an order of magnitude. However, the interaction picture methodology requires a time-dependent simulation method, rendering it incompatible with some simulation techniques such as qubitization (although recent work has shown that qubitization can be made to work for restricted families of time-dependent Hamiltonians, existing results preclude the interaction picture~\cite{watkins2022time}). For the time-independent case, the symmetry shift technique here introduced can be employed, which yields a significant decrease of the 1-norm when compared with commonly used techniques \cite{femoco_df} and can be included in a tensor hypercontraction framework \cite{THC}. Even though we only used symmetry shifts that can be written as functions of occupation number operators such as $\hat N_e$ and $\hat S_z$, the results presented here provide an incentive for exploring more symmetries with different structures, such as $\hat S^2$ and molecular point group symmetries. Finally, we have found tight bounds for the 1-norm of an arbitrary \gls{LCU}, with the most efficient decomposition shown here being close to the lower bound. \\

Overall, we presented different methodologies for reducing the \gls{LCU} 1-norm cost of molecular Hamiltonians, with significant improvements that can be applied both for time-independent \cite{qubitization_df} and time-dependent \cite{LCU,interaction} simulation methods. We expect our results to be of use for any methodology that uses \gls{LCU} decomposition of molecular Hamiltonians.

\section*{Acknowledgements}
I.L. is grateful to Tzu-Ching Yen for helpful discussions and to Luis Martínez-Martínez, Joshua T. Cantin, and Seonghoon Choi for their comments on the manuscript, and is supported by the funding of the Anoush Khoshkish Graduate Research Scholarship in Chemistry. A.F.I. acknowledges financial support from the Google Quantum Research Program, Zapata Computing, and the Natural Sciences and Engineering Research Council of Canada.  N.W. acknowledges funding from the Google Quantum Research Program, the Natural Sciences and Engineering Research Council of Canada and N.W.’s theoretical work on this project was supported by the U.S. Department of Energy, Office of Science, National Quantum Information Science Research Centers, Co-Design
Center for Quantum Advantage under contract number
DE-SC0012704.

\appendix

\section{Proof of the lowest 1-norm theorem}
\label{app:proof}
\bound*
The construction of an \gls{LCU} that achieves the lower bound is given in Sec.~\ref{subsec:theorem} of the main text. We now proceed to show, by contradiction, that no \gls{LCU} with a lower 1-norm can be found.

\begin{proof}[Proof of~Theorem~\ref{thm:ck}] 
Let us assume that it is possible to construct $\hat H = \sum_k u_k \hat U_k$ such that $\sum_k |u_k| < \Delta E/2$.  Denoting $\ket{\psi_{\max}}(\ket{\psi_{\min}})$ the maximum(minimum) eigenvectors, we have
\begin{align}
 \bra{\psi_{\max}} \sum_k u_k \hat U_k \ket{\psi_{\max}} = E_{\max} \nonumber\\
    \bra{\psi_{\min}} \sum_k u_k \hat U_k \ket{\psi_{\min}} = E_{\min}.
\end{align}
We therefore have that $\Delta E/2$ can be bounded above by
\begin{align}
     \frac{\Delta E}{2} = &\frac{1}{2}\Big(\bra{\psi_{\max}} \sum_k u_k \hat U_k \ket{\psi_{\max}} -\bra{\psi_{\min}} \sum_k u_k \hat U_k \ket{\psi_{\min}}\Big) \nonumber\\
     & \quad \le \sum_k |u_k| < \frac{\Delta E}{2},
\end{align}
where we have used the triangle inequality and the fact that $|\bra{\psi} \hat U \ket{\psi}|\leq 1$ for any unitary $\hat U$ and wavefunction $\ket{\psi}$.
This is a contradiction, thus $\sum_k |u_k| \ge \Delta E/2$ as claimed.

\end{proof}

\section{One-electron term transformations} 
\label{app:tensors}
Due to idempotency of occupation operator terms, $(\hat E^{i\sigma}_{i\sigma})^2 = \hat E^{i\sigma}_{i\sigma}$,
one-electron terms can be transformed to two-electron terms and vice versa.  
This section shows how these transformations can be done for general two-electron operators.

\subsection{Transforming to two-electron terms}
We start by showing how to combine one- and two-electron terms into a sum of two-electron terms. An arbitrary operator with spin-orbit symmetry can be written as
\begin{align}
    \hat O &= \hat O_1 + \hat O_2 \\
    &= \sum_{\sigma}\sum_{ij}o_{ij}\hat E^{i\sigma}_{j\sigma} + \sum_{\sigma\sigma'}\sum_{ijkl}o_{ijkl}\hat E^{i\sigma}_{j\sigma}E^{k\sigma'}_{l\sigma'}. \label{eq:gen_op}
\end{align}
We note that each $o_{ij}$ is multiplied by two excitation operators $\hat E^{i\sigma}_{j\sigma}$ with $\sigma \in (\alpha,\beta)$, while each $o_{ijkl}$ is multiplied by four operators ($\alpha\alpha,\alpha\beta,\beta\alpha,\beta\beta$). Because of this, combining both operators requires to change from a (spacial) orbital representation into the spin-orbital one; the combined two-electron operator will have different components for ($\alpha\alpha,\beta\beta$) and ($\alpha\beta,\beta\alpha$). We now transform the one-electron term into a two-electron operator, noting that several different transformations can be done here. We use the one that maintains a symmetric two-electron tensor since this property makes computational manipulations more efficient:
\begin{align}
    \hat O_1 &= \hat U \sum_{\sigma}\sum_{i} \mu_i \hat n_{i\sigma} \hat U^\dagger \\
    &= \hat U \sum_{\sigma}\sum_{i} \mu_i \hat n_{i\sigma} \hat n_{i\sigma} \hat U^\dagger \\
    &= \sum_{\sigma} \sum_{ijkl} \tilde o_{ijkl} \hat E^{i\sigma}_{j\sigma} \hat E^{k\sigma}_{l\sigma},
\end{align}
with 
\begin{equation}
    \tilde o_{ijkl} = \sum_m \mu_m U_{im} U_{jm} U_{km} U_{lm},
\end{equation}
for $\hat U \equiv \sum_\sigma \sum_{ij} U_{ij} \hat E^{i\sigma}_{j\sigma}$. The combined two-electron tensor is $o_{pqrs}$, where
\begin{align}
    \hat O &= \sum_\sigma \sum_{ijkl} \Big(\tilde o_{ijkl} + o_{ijkl}\Big) \hat E^{i\sigma}_{j\sigma} \hat E^{k\sigma}_{l\sigma} + \sum_{\sigma\neq\sigma'}\sum_{ijkl} o_{ijkl}\hat E^{i\sigma}_{j\sigma}E^{k\sigma'}_{l\sigma'} \\
    &\equiv \sum_{pqrs} o_{pqrs} \hat E^p_q \hat E^r_s.
\end{align}

\subsection{Separating one-electron terms}
Here we show how one-electron terms can be separated from the two-electron tensor, which is used to connect the fermionic and qubit representations. This is necessary for writing the 1-norm of qubit-based methods as a function of the molecular integrals, and also for a more efficient sorted-insertion algorithm that does not require any fermion to qubit mappings. This methodology is introduced in Ref.~\citenum{majorana_l1}, only the final result and key steps are presented here. We start from a general two-electron operator with spin-symmetry, as seen in Eq.\eqref{eq:gen_op}. For the next step, we switch to the Majorana representation, where fermionic operators are mapped into Majorana operators as
\begin{align}
    \hat \gamma_{j\sigma,0} &= \hat a_{j\sigma} + \hat a_{j\sigma}^\dagger \\
    \hat \gamma_{j\sigma,1} &= -i(\hat a_{j\sigma} - \hat a_{j\sigma}^\dagger),
\end{align}
which have the well-known algebraic relations:
\begin{align}
    \{\hat \gamma_{p,n}\hat \gamma_{q,m}\} &= 2\delta_{pq}\delta_{nm}\hat 1 \\
    \hat\gamma_{p,n}^\dagger &= \hat\gamma_{p,n} \\
    \hat\gamma_{p,n}^2 &= \hat 1,
\end{align}
for $n,m = 0,1$. By transforming the fermionic operators into Majoranas, after proper manipulations, $\hat O$ can be written as
\begin{align}
    \hat O &= \Big(\sum_i o_{ii} + \frac{1}{2}\sum_{ij}o_{ijji} + \sum_{ij} o_{iijj} \Big)\hat 1 \nonumber \\
    &\ \ + \frac{i}{2}\sum_{\sigma}\sum_{ij}\Big(o_{ij} + 2\sum_k o_{ijkk} \Big)\hat\gamma_{i\sigma,0}\hat\gamma_{j\sigma,1} \nonumber \\
    &\ \ - \frac{1}{4} \sum_{\sigma\neq\sigma'}\sum_{ijkl} o_{ijkl} \hat\gamma_{i\sigma,0}\hat\gamma_{j\sigma,1}\hat\gamma_{k\sigma',0}\hat\gamma_{l\sigma',1} \nonumber \\
    &\ \ - \frac{1}{2}\sum_\sigma\sum_{i>k,l>j}(o_{ijkl} - o_{ilkj})\hat\gamma_{i\sigma,0}\hat\gamma_{j\sigma,1}\hat\gamma_{k\sigma,0}\hat\gamma_{l\sigma,1}.
\end{align}
In this equation, all terms with a product of 2 Majorana operators correspond to the one-electron terms, while terms with 4 Majorana operators correspond to two-electron terms. This representation thus allows for the analytical separation of the one-electron terms that are present in the two-electron tensor. Finally, we note that each resulting Majorana product will map to an individual Pauli product when using fermion to qubit mappings such as Jordan-Wigner or Bravyi-Kitaev.

\section{Finding the optimal symmetry shift for 1-norm minimization}
In this section we present procedures for finding the coefficient vector associated with the symmetry shift procedure in Sec.~\ref{subsec:symmetries}, both for the one- and two-electron cases. 

\subsection{Two-electron symmetry reduction} \label{app:sym_L1}
For finding the symmetry shift of two-electron operators, we here present a linear programming routine that minimizes the 1-norm and is computationally efficient. Since the 1-norm of an arbitrary two-electron operator depends on the unitarization method, we minimize an approximate 1-norm instead for computational efficiency, corresponding to $\sum_{ij} |\tilde g_{iijj}|$ for an operator as seen in Eq.\eqref{eq:mol_ham}. \\

Our 1-norm minimization routine can be written for arbitrary occupation number polynomial symmetries corresponding to $\hat H - \sum_u s_u \hat S_u$, which can be written as
\begin{equation}
\label{eq:sr-l1norm}
\min_{\overrightarrow{s}} \sum_{\nu=1}^{|C|} |\lambda_{\nu} - \sum_{u=1}^{|\mathbb S|} s_u \tau^{(u)}_\nu|.
\end{equation}
In the above equation, vector $\overrightarrow{\lambda}$ collects the diagonal elements of the two-electron tensor that appear in the array of Cartan operators $C = \{ \hat{n}_{p} \hat{n}_{q} \mid p \geq q\} \equiv \{\hat C_\nu \mid \nu=1,...,|C|\}$. The vectors we use for this notation run over $p\geq q$ in the same order as $C$, with the number of elements $|C| = n(n+1)/2$. This also defines the $\vec \tau^{(u)}$ vectors, which represent the symmetry operators: $\hat S_u = \sum_{\nu=1}^{|C|} \tau^{(u)}_{pq} \hat C_\nu$. 
As outlined in the main text, the only symmetry that we consider for the two-electron component is $\hat N_e^2$, which corresponds to $\tau_{pq} = 1,\ \forall p,q$. However, we outline the linear programming routine for arbitrary symmetry operator pools.

We used a linear programming technique to find $\overrightarrow{s}$ efficiently while directly minimizing the 1-norm.
To this end, we need to transform the sum of absolute values in the 1-norm cost function into a sum of linear parameters. 
We achieve this by introducing a new set of auxiliary parameters equivalent to the absolute value functions that appear in the 1-norm. The minimization procedure is subject to constraints that satisfy the absolute value conditions.
For the 1-norm linearization, let us begin with introducing the auxiliary vector $\overrightarrow{t}$ and re-write the objective function as 

\begin{equation}
\begin{aligned}
& \min_{\vec s} \sum_{\nu} t_{\nu}(\vec s) \\
& \text{subject to}
& t_{\nu}(\vec s) = \sum_{\nu}|\lambda_{\nu} - \sum_u s_u \tau^{(u)}_\nu | .  
\end{aligned}
\end{equation}
The absolute value constraints can be written as 

\begin{equation}
 \sum_u s_u\tau^{(u)}_\nu - \lambda_{\nu}  \leq  t_{\nu}
\end{equation}
and
\begin{equation}
-\sum_u s_u \tau^{(u)}_\nu + \lambda_{\nu} \leq t_{\nu} 
\end{equation}
We can now write the final form for the 1-norm minimization problem:
\begin{equation}
\begin{aligned}
& \min_{ \{\vec s,\vec t \} } \sum_{\nu} t_{\nu} \\
& \text{subject to}
& & \begin{pmatrix}
\underline \tau & -\underline 1 \\ 
-\underline \tau & -\underline 1
\end{pmatrix} \begin{pmatrix}
\vec s\\ 
\vec t
\end{pmatrix} \leq
\begin{pmatrix}
 \vec \lambda \\ 
 -\vec \lambda
\end{pmatrix},
\end{aligned}
\end{equation}
where we have defined the matrix $\underline\tau$ such that the product $(\underline\tau \times \vec s)_\nu \equiv \sum_u s_u\tau_\nu^{(u)}$, and the inequality runs over each index $\nu \in \{1,...,|C|\}$. The above minimization problem is linear programmable and lets us efficiently find the global minimum of the 1-norm of the objective function. Moreover, the $\underline\tau$ matrix helps to avoid storing in memory the full two-electron tensors of symmetry operators, which reduces the memory demand significantly. We used the highs global minimization algorithm~\cite{highs} for our numerical calculations.

\subsection{One-electron symmetry reduction} \label{app:OBsym}
Here we review the method for finding the optimal symmetry shift minimizing the 1-norm of the one-electron term $\hat H_1'(s_2)$ obtained after including the one-body modification from $s_2\hat N_e^2$ to $\hat H_1'$ [\eq{eq:Tmod}]. This method optimizes a real number $s_1$, such that the resulting 1-norm of the operator $\hat H_1''(s_1,s_2) \equiv \hat H_1'(s_2) - s_1 \hat N_e$ is minimized. Since $\hat N_e$ commutes with orbital rotations, 
\begin{align}
    \hat H_1''(s_1,s_2) &= \hat U^\dagger \left(\sum_{i\sigma} \mu_i(s_2) \hat n_{i\sigma}\right) \hat U - s_1 \hat N_e \\
    &= \hat U^\dagger \left ( \sum_{i\sigma} (\mu_i(s_2) - s_1) \hat n_{i\sigma} \right) \hat U,
\end{align}
therefore, minimizing the 1-norm of $\hat H_1''(s_1,s_2)$ is equivalent to minimizing $\sum_i |\mu_i(s_2) - s_1|$, which can be done efficiently using linear programming as outlined for the two-body symmetry shift routine.

\section{Molecular geometries and computational details}
In this section we write down all details necessary for numerical reproducibility. Our code is available at \href{https://github.com/iloaiza/MAMBO}{https://github.com/iloaiza/MAMBO}. \\

For the CSA decompositions of the two-electron tensor, the cost function was chosen as the 2-norm of $\Delta \hat H \equiv \sum_{\sigma\sigma'}\sum_{ijkl} b_{ijkl}\hat E^{i\sigma}_{j\sigma}\hat E^{k\sigma'}_{l\sigma'}$:
\begin{equation}
    ||\Delta\hat H||_2 \equiv \sum_{ijkl} |b_{ijkl}|^2,
\end{equation}
considering the decomposition finished when this norm is below a tolerance of $1\times 10^{-6}$. All non-linear optimizations where done using the Julia Optim.jl package \cite{optim}, using the BFGS algorithm \cite{bfgs} with the default tolerance. Linear programming routines were done using the Julia JuMP package \cite{jump} with the HiGHS optimizer \cite{highs}.

\subsection{Molecular Hamiltonians}
All molecular Hamiltonians were generated using the PySCF package \cite{pyscf1,pyscf2,pyscf3} and the Openfermion library \cite{openfermion}, using a minimal STO-3G basis \cite{sto3g,szabo} and the Jordan-Wigner transformation \cite{jordan_wigner}. The nuclear geometries for the Hamiltonians are:
\begin{itemize}
    \item R(H -- H) = $1\textrm{\AA}$ for H$_2$
    \item R(Li -- H) = $1\textrm{\AA}$ for LiH
    \item R(Be -- H) = $1\textrm{\AA}$ with a collinear atomic arrangement for BeH$_2$
    \item R(O -- H) = $1\textrm{\AA}$ with angle $\angle$HOH = $107.6^\circ$ for H$_2$O
    \item R(N -- H) = $1\textrm{\AA}$ with $\angle$HNH = $107^\circ$ for NH$_3$
\end{itemize}

\bibliography{main}

\end{document}